\documentstyle[12pt]{article}
\newcommand{\la}{\label}

\newcommand{\be}{\begin{equation}}
\newcommand{\ee}{\end{equation}}
\newcommand{\ba}{\begin{eqnarray}}
\newcommand{\ea}{\end{eqnarray}}
\newcommand{\bastar}{\begin{eqnarray*}}
\newcommand{\eastar}{\end{eqnarray*}}

\begin{document}
\begin{titlepage}
 
 
\vskip 0.4truecm
 
\begin{center}
{ 
\bf \large \bf KNOTS AND PARTICLES \\
}
\end{center}
 
\vskip 2.8cm
 
\begin{center}
{\bf L. Faddeev$^{* \sharp}$ } {\bf \ and \ } {\bf 
Antti J. Niemi$^{** \sharp}$ } \\
\vskip 0.4cm

{\it $^*$St.Petersburg Branch of Steklov Mathematical
Institute \\
Russian Academy  of Sciences, Fontanka 27 , St.Petersburg, 
Russia$^{\ddagger}$ } \\

\vskip 0.4cm

{\it $^{**}$Department of Theoretical Physics,
Uppsala University \\
P.O. Box 803, S-75108, Uppsala, Sweden$^{\ddagger}$ } \\

\vskip 0.2cm
and \\
\vskip 0.2cm

{\it $^{\sharp}$Research Institute for Theoretical Physics \\
P.O. Box 9, FIN-00014 University of Helsinki, Finland} \\

\end{center}

\vskip 0.7cm
\rm
\noindent
Using methods of high performance computing, 
we have found indications 
that knotlike structures appear 
as stable finite energy solitons 
in a realistic 3+1 dimensional model. We 
have explicitly simulated the unknot and trefoil 
configurations, and our results suggest that all 
torus knots appear as solitons.
Our observations open new theoretical possibilities in
scenarios where stringlike structures appear,
including physics of fundamental interactions and 
early universe cosmology. In nematic
liquid crystals and $^{3}$He superfluids such 
knotted solitons might actually be observed.
 
\vfill
 
\begin{flushleft}
\rule{5.1 in}{.007 in} \\
$^{\ddagger}$  \small permanent address \\ \vskip 0.2cm
$^{*}$ \small supported by Russian Academy of Sciences
and the Academy of Finland \\ \vskip 0.2cm 
$^{**}$ \small Supported by G{\"o}ran Gustafsson
Foundation for Science and Medicine \\
\hskip 0.3cm and by NFR Grant F-AA/FU 06821-308
\\ \vskip 0.3cm
$^{*}$ \hskip 0.2cm {\small  E-mail: \scriptsize
\bf FADDEEV@PDMI.RAS.RU and FADDEEV@PHCU.HELSINKI.FI } \\
$^{**}$  {\small E-mail: \scriptsize
\bf ANTTI.NIEMI@TEORFYS.UU.SE}  \\
\end{flushleft}

\end{titlepage}
In 1867 Lord Kelvin \cite{kelvin} proposed
that atoms, which at the time were considered
as elementary particles, are knotted vortex 
tubes in ether. For about 
20 years his theory was taken seriously, and 
motivated an extensive study of 
knots. The results obtained at the time
by Tait \cite{tait} remain a 
classic contribution to mathematical
knot theory \cite{ati}. More recently the idea that 
elementary particles can be identified
with topologically distinct knots 
has been advanced in particular by Jehle \cite{jehle}.
 
Today it is commonly accepted that
fundamental interactions are
described by stringlike structures \cite{green},
with different elementary particles corresponding
to the vibrational excitations of a fundamental
string. Even though there are hints of
connections between modern string theory 
and knot theory \cite{witten},
knotlike structures have
not yet been of much 
significance.

There are also a number of other scenarios where
knotted structures may enter Physics 
\cite{ati}, \cite{rebbi}.
These include models in statistical physics,
QCD strings that confine
quarks inside nuclear particles,
cosmic strings that are expected
to be responsible for early universe structure formation,
and approaches to quantum gravity where knots
are supposed to determine gauge invariant observables.
Stringlike vortices appear in type-II superconductors 
where they confine magnetic fields within the cores of 
vortex-like structures, and  similar phenomena 
are also present in superfluid $^4$He.
Recent experiments with nematic liquid crystals
\cite{bowick} and $^3$He superfluids \cite{lou} 
have also revealed interesting vortex structures 
that can be described by theoretical methods 
which are adopted from cosmic string models. 
Finally, physics of knots
is rapidly becoming an important
part of molecular biology, 
where entanglement of a DNA chain interferes with 
vital life processes
of replication, transcription and 
recombination \cite{dna}.

Thus far the physics of knotlike structures 
has been investigated 
sparsely. This is largely
due to a lack of dynamical principles that enable 
the construction of stable knots. One 
needs a theoretical model where knots emerge
as solitons, {\it i.e.} as stable finite energy 
solutions to the pertinent nonlinear field equations.

The literature on solitons is enormous,
and there are several extensive reviews \cite{rebbi}, 
\cite{fadde1}, \cite{sany}. Until 
now the activity has mainly concentrated
on 1+1 dimensions with the notable exceptions of
the 2+1 dimensional vortex and nonlinear $\sigma$-model
soliton, 
and skyrmeons and 'tHooft-Polyakov monopoles in 3+1 dimensions.
These are all pointlike configurations, 
and can not be directly associated with knotlike structures.

When embedded in three dimensions, a pointlike 
two dimensional soliton becomes a line vortex.
For a finite energy its length must
be finite which is possible if its core 
forms a knot. In 1975  one of us \cite{fadde2} 
proposed that closed, knotted vortices could be
constructed in a definite dynamical model. 
The explicit solution suggested 
in \cite{fadde2} is a closed torus-like vortex
ring, twisted once around its core 
before joining the ends which ensures 
stability against shrinking. This closed
vortex corresponds to the {\it unknot} (see figure 4)
which is the simplest possible knotlike structure.
However, despite numerous attempts
no such stable configuration has been constructed, neither
by explicit analysis nor by a numerical investigation.

\vskip 0.3cm
Here we shall report on our
work to construct knotlike vortices in the
model introduced in \cite{fadde2}.
By employing numerical algorithms in
powerful computers we have been able
to find strong evidence for the existence
of the unknot vortex.
In addition we have found indications 
supporting the existence of a  {\it trefoil } vortex
(see figure 5).
A trefoil is the simplest example of 
torus-knots, which are obtained by winding around
a torus in both directions \cite{ati}.
Our results suggest that in fact all torus
knots should appear as vortex solitons in the 
model proposed in \cite{fadde2}. 
This model describes the 3+1 dimensional
dynamics of a three component vector
$\vec{\bf n}( {\bf
x},\tau) $ with unit length, $ \vec{\bf n} 
\cdot \vec { \bf n} = 1 $. 
Such a vector field is a typical degree of freedom
in the nonlinear $\sigma$-model, a prototype relativistic
quantum field theory. It also appears  
as an order parameter in 
the Heisenberg ferromagnet. A unit vector field is
also present in models of nematic liquid crystal
where it characterizes the average 
direction of the rod, and
in $^3$He-A superfluid where it 
determines the spin projection direction for a 
Cooper pair. Indeed, the model proposed
in \cite{fadde2} is quite universal, and we expect
our results to have a large number of 
applications.

In order that $\vec {\bf n}( \vec {\bf x} ) $ 
describes a localized stationary knot, it must go to
a constant vector $\vec{ \bf n}(\vec {\bf x}) \to 
\vec{\bf n}_0$ at large distances.
Consequently $\vec {\bf n}( \vec {\bf x} )$
defines a mapping from the compactified $R^3 \sim S^3 
\to S^2$. Such mappings fall into nontrivial homotopy classes
$\pi_3(S^2) \simeq Z$ and can be characterized by
the Hopf invariant \cite{ati}. For this we introduce
the two-form $ F =  ( d \vec{\bf n} \wedge 
d \vec{\bf n} , \vec{\bf n} )$ on the target $S^2$.
Its preimage $F_\star $ on $S^3$ is 
exact, $ F_\star = d A_\star$, 
and the Hopf invariant $Q_H$ is given by
\be
Q_H \ = \ \int\limits_{R^3} F \wedge A 
\la{hopf}
\ee

The Hamiltonian proposed in \cite{fadde2} is 
\be
H \ = \ E_2 \ + E_4 \ = \
\int d^3 x \ g^2 ( \partial_\mu 
\vec {\bf n})^2 + 
e^2 F^2
\la{ham}
\ee
This is the most general three dimensional Hamiltonian 
that admits a relativistically invariant extension in
3+1 dimensions and involves only terms with no more than four 
derivatives. It can be related to the $SU(2)$ Skyrme model when
restricted to a sphere $S^2 \in SU(2)$, but its topological 
features are different. In particular,
the existence of nontrivial knotted vortex solutions  
in (\ref{ham}) is strongly suggested by the lower bound 
$ H  \geq  c \cdot | Q_H |^{\frac{3}{4}}$ \cite{vak}.

The first term $E_2$ determines the standard 
nonlinear O(3) $\sigma$-model which admits 
static solitons in two dimensions.
But in three dimensions a scaling $ \vec{\bf x} 
\to \rho \vec{\bf x}$ reveals that stable 
stationary solutions are possible 
only if $E_4$ is also present. Indeed,
under this scaling $ E_2 \to \rho E_2$ but
$ E_4 \to \rho^{-1} E_4 $
from which we conclude that in three dimensions
finite energy solutions obey
the virial theorem,
\be
E_2 \ = \ E_4
\la{virial}
\ee

Several articles have been devoted  
for analyzing the general properties of
the unknot vortex in the model (\ref{ham}) \cite{sany}.
However, to our knowledge there 
have been no real attempts to find an 
actual solution. This is due to the fact,
that even in the simplest case of an unknot
the separation
of variables eliminates only one of the three
space coordinates, leaving 
numerical methods as the sole alternative
for finding a solution. With
the recent, enormous progress in supercomputing
techniques serious attempts are
finally becoming realistic. 

The Euler-Lagrange equations for 
(\ref{ham}) determine a highly 
nonlinear elliptic system,
and a direct numerical approach 
appears to be complicated.
Instead, we consider the parabolic equation
\be
\frac{d \phi_a }{ dt } \ = \ - \ \frac{ \delta 
H (\phi) }{\delta \phi_a }
\la{eom}
\ee
where $\phi_a$ denotes a generic
dynamical variable in (\ref{ham}). 
The $t$-bounded solutions of (\ref{eom}) 
connect the
critical points of $H$ by flowing away from 
an unstable critical point at $t \to -\infty$
towards a stable critical point at $t \to + \infty$.
From this we expect that by defining a suitable
initial configuration at $t=0$, for large $t \gg 0$
we flow towards a stable vortex solution of 
the original stationary equation.

We parametrize our 
unit vector by $\vec{\bf n} = (\cos \varphi 
\sin \theta, \sin \varphi \sin \theta , \cos \theta) $
through stereographic coordinates
$ \varphi =  - \arctan ( \frac{V}{U} ) $ and
$ \theta =  2 \arctan \sqrt{ U^2 + V^2 } $. 
In terms of these variables the Hamiltonian (\ref{ham})
becomes
\be
H   =  \int d^3x \ { 
4 g^2 \over (1 + U^2 + V^2)^2 } ( \partial_\mu U ^2 
+ \partial_\mu V^2 )
+ 
{ 16 e^2 \over (1 + U^2 + V^2)^4 } 
( \partial_\mu U \partial_\nu V
\ - \ \partial_\nu U \partial_\mu 
V )^2
\la{ham2}
\ee

By a global $SO(3)$ rotation we can
select our asymptotic vacuum vector $\vec{\bf n}_0$
so that outside of the vortex it points to 
the positive-$z$ direction. 
This means that outside of the
vortex we are near the north pole where
$\theta ( \vec{\bf x } ) \approx 0$, while 
the core corresponds to the 
south pole $\theta( \vec{\bf x }_c ) = \pi$.
The internal structure of a vortex can then
be investigated
by cutting it once with a plane at 
right angle to its core.     
This cross sectional plane is topologically 
identical to a sphere $S^2$: At the core we
have $\theta( \vec{\bf x}_c) = \pi$ corresponding
to the south pole and
outside of the
vortex on the plane we have $\theta (\vec{\bf x }) 
\approx 0$ corresponding
to the north pole. Furthermore,  
$\varphi (\vec{\bf x})$ increases (or decreases, depending
on orientation)
by $2\pi$ when we go around the core once on the cross 
sectional plane.\footnotemark\footnotetext{More generally,
$\varphi$ is defined modulo $2\pi n$ where $n$ is an integer.
Here we only consider the simplest case with $n = \pm 1$.}
For $U$and $V$ this means, that we identify
them as local coordinates on the Riemann sphere in a patch
that contains the north pole $U (\vec{\bf x }) = 
V (\vec{\bf x }) = 0$.

In figures 1 and 2 we have drawn
$\theta (\vec{\bf x }) $ and $\varphi (\vec{\bf x})$
respectively for our unknot vortex. The general
structure is clearly visible
in these pictures. 

In the present model there are two symmetries
that relate a given vortex 
solution with Hopf invariant $Q_H$ to an 
antivortex solution with
equal energy but opposite Hopf invariant $-Q_H$.
The first symmetry emerges when we
change orientation $\varphi \to -\varphi$ on the cross 
sectional plane.
The second symmetry interchanges the
north pole with the south pole on the cross sectional
plane, {\it i.e.} maps the core to
$\theta (\vec{\bf x}_c) = 0$ and the region outside
of the vortex to $\theta (\vec{\bf x }) 
\approx \pi$. In terms of 
the $U(\vec{\bf x })$ and
$ V(\vec{\bf x })$ variables this means that
on each cross sectional plane 
the Hamiltonian (\ref{ham2}) must be invariant under
the inversion $ Z (\vec{\bf x }) =  U(\vec{\bf x }) + i V
(\vec{\bf x })  \to Z^{-1} (\vec{\bf x }) $. 
This implies in particular, that the energy
density $H(\vec{\bf x })$ must be 
concentrated in a relatively
narrow tubular neighborhood around 
the core of the vortex.

In figure 3 we have depicted the energy density on a
cross sectional plane for the unknot vortex. The concentration
of energy density in a tubular manner 
around the core is clearly visible.

In order to specify the initial condition
in (\ref{eom}) we need 
a knot configuration which is topologically
equivalent
to the desired vortex solution. We specify this configuration
by first introducing a parametrization $\vec{\bf x}(\lambda)$ 
for the center of the knot in $R^3$, and then
for each $\lambda$ use the Serret-Frenet 
equations to define local
coordinates  $U(\vec{\bf x} ; \lambda)$ 
and $V(\vec{\bf x} ; \lambda)$
on the cross sectional planes.

The choice of initial parametrization
$\vec {\bf x}(\lambda)$ for the core
introduces a scale which may be quite
different from the one specified
by the coupling constant $g$ in (\ref{ham2}). 
In order to enhance convergence, we
adopt a simple renormalization procedure
by promoting $g$ to a time dependent
function $g(t)$. We define this 
time dependence
by demanding that at each value of $t$ the virial
theorem (\ref{virial}) must be obeyed. Since
a vortex solution 
obeys (\ref{virial}), this means that $g(t)$
flows towards a fixed point value $
g(t) \ \to \ g^{\star} $.

\vskip 0.3cm
We have performed our numerical simulations
using version 5.3 of G. Sewell's PDE2D finite element
algorithm  \cite{sewell}.
During the early phase of our simulations we
have used the
initial knot configuration
to determine the boundary conditions
on the finite element mesh.
However, since
the exact boundary conditions for the desired vortex
are {\it a priori} unknown, after several
time steps we have decreased the size of the
finite element mesh 
and used the pertinent
simulated configuration to determine the boundary conditions
in the new mesh. By starting from a sufficiently
large initial mesh, we then eventually obtain
a submesh with boundary conditions that are close
to those of the actual solution.

For our simulation we have used
two Silicon Graphics Power Challenge computers
with R8000 processors equipped with
1GB {\it resp.} 2GB of internal memory.

\vskip 0.3cm
For the unknot vortex ($Q_H = \pm 1$), 
the equations of motion
can be simplified using axial symmetry. 
We select the symmetry axis 
to coincide with the $z$-axis in $R^3$, and
introduce cylindrical coordinates $
r, \phi, z$. With the Ansatz $
\varphi (\vec{\bf x})  = 
\varphi(r, z) +   \phi $ and 
$\theta (\vec{\bf x } )  =  \theta (r , z ) $
the $\phi$-coordinate separates, and 
we obtain a two dimensional equation for $U(r, z)$ 
and $V(r,z)$. That 
such a separation of variables is possible
follows directly from the $SO(2) \times SO(2)$
symmetry of the unknot configuration.
The ensuing equations are
defined on the half-plane $r \geq 0$, $z \in 
( -\infty, \infty)$, which at each $\phi$
determines our cross sectional plane.
Besides the inversion invariance, the equations
of motion are now also invariant under $z \to -z$.
Since the unknot solution is even in $z$, this halves the
CPU time in our simulation.

In figures 1-4 we describe an unknot vortex
which is a result of over 50h of CPU time with
each time step taking about 9 minutes
on a 9600 triangle mesh. 
The mesh has been selected so that it is 
more dense at the boundaries and
near the core of the vortex. 
In our simulation we find impressive convergence,
allowing us to increase the time step by up to
8 orders of magnitude while keeping the relative
change in energy intact. The Hopf invariant
is also very stable, and we have identified
the final configuration on the basis that the
numerically computed Hopf invariant has 
a slight local maximum with $Q_H = 0.999996 \ ...$

\vskip 0.3cm
The trefoil vortex ($Q_H = \pm 3$)
described in figure 5 is a result
of almost 200h in CPU time on a $21^3$ cubic 
finite element lattice with tri-cubic 
Hermite basis functions. Consequently 
the number of nodes is about the same
as in the case of unknot, but due to a lack
of any obvious reflection symmetry each time iteration
now takes about 20 minutes of CPU time.
The yellow center in figure 5 corresponds to
the core $\vec{\bf x}(\lambda)$ of our initial 
configuration, which has been determined using the 
energy concept developed by J. Simon 
\cite{simon}. The points in figure 5 have been 
evaluated using the piecewise cubic polynomial
approximation obtained from the finite
element algorithm. 
As in figure 4, the picture 
describes
a volume where the
energy density inside the vortex essentially
vanishes, {\it i.e.} it 
can be viewed as an extended core
of our trefoil vortex. 

We expect that the nonhomogeneity
in figure 5 
reflects the underlying lattice structure that
we have used in our
simulation.
Indeed, a $21^3$ lattice is obviously too rough to
describe our trefoil solution adequately, but 
unfortunately
we do not have access to a computer that would
allow us to use an essentially larger lattice.
Nevertheless, we have found definite numerical 
stability in the sense that the final 
configuration in figure 5 has remained essentially 
intact under a large number of iterations. 
We view this stability as a strong evidence that
we indeed have convergence towards a trefoil
vortex solution.
 
\vskip 0.4cm
In conclusion, we have performed numerical
simulations with a high-performance computer
to investigate knotted vortex solutions
in the model introduced in \cite{fadde2}. 
By investigating the unknot
and trefoil vortices, we
have found strong evidence that torus knots 
indeed appear as solitons.
For the unknot, we have found very impressive
convergence and our results for the trefoil are
also quite encouraging. However, since the
computers we can access do not allow us to
effectively study dense
three dimensional lattices,  our simulation of 
the trefoil is still tentative.

We expect that our results will have numerous
important
applications. In particular, 
since the order parameter in nematic liquid
crystal and $^3$He superfluid 
involves a unit three vector, an experimental
investigation of vortices in these materials
should reveal the existence
of stable knotlike structures. We also expect that
an extension of our work to spontaneously
broken Yang-Mills-Higgs theories
where stable knotlike vortices can
not be excluded by scaling arguments, should have 
interesting physical implications in particular
to early universe cosmology.

\vskip 1.0cm
We wish to thank  J.Hietarinta, T.K\"arki,   
A.P.Niemi, K.Palo, J.Pitk\"aranta, J.Rahola, D.Riska, 
R.Scharein, 
G.Sewell,  O.Tirkkonen and J.Tulkki for helpful discussions
and valuable suggestions.
We are particularly indebted to Sami Virtanen for his help
with visualization and simulating the initial Ansatz for
the trefoil, and Matti Gr\"ohn for helping us
with visualization. We are grateful to the Center
for Scientific Computing in Espoo, Finland for 
providing us with an access to their line of
Silicon Graphics Power Challenge computers.

\vfill\eject

\vfill\eject
\begin{flushleft}
{\bf Figure Caption}
\end{flushleft}
\vskip 1.0cm

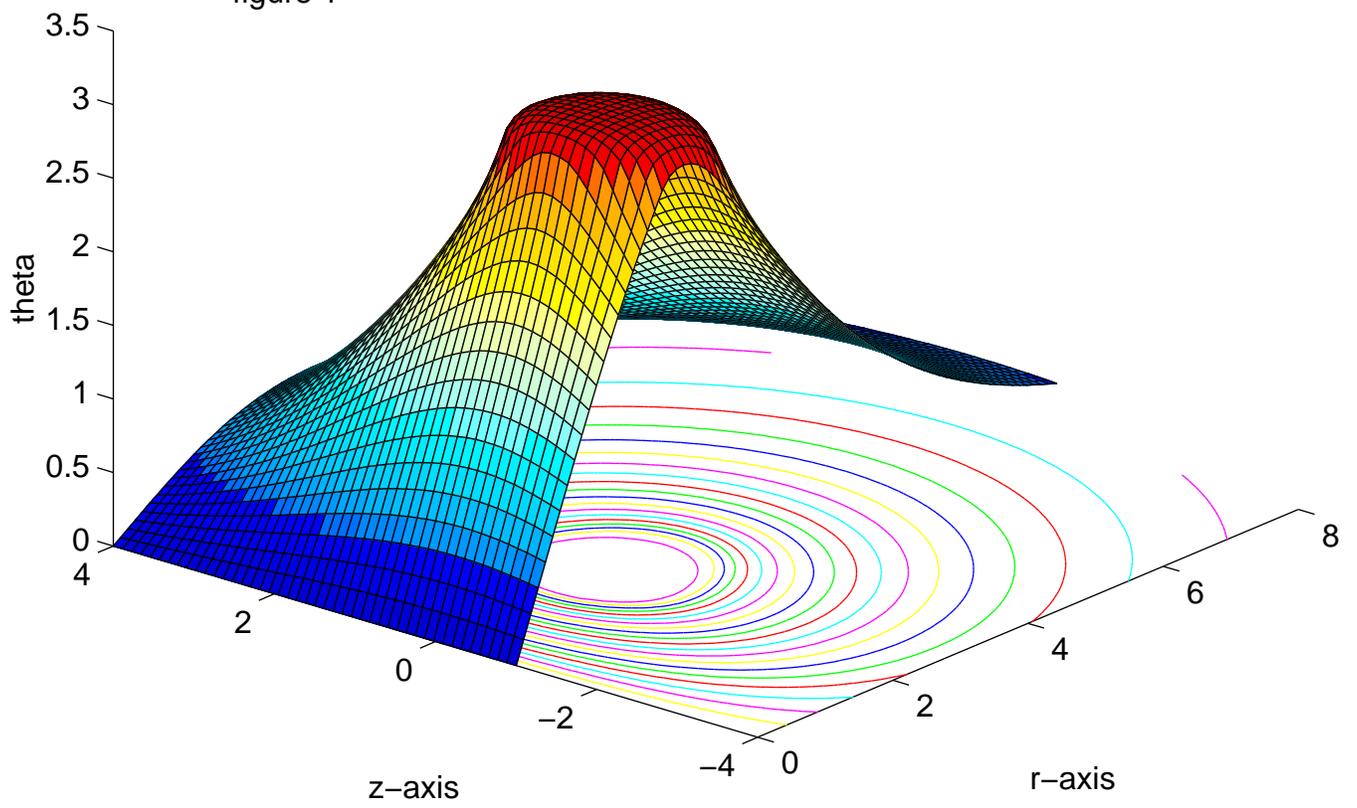
{\bf Figure 1:} A combined surface and contour 
plot of $\theta(r,z)$ for an unknot vortex
with $g^\star \approx 0.24, \ e = 1 $ 
in cylindrical coordinates $(r, \phi, z)$
and on a cross sectional plane with generic $\phi$.
The configuration is $z \to -z$ symmetric.

\vskip 0.4cm
{\bf Figure 2:} A combined surface and contour 
plot of $\varphi(r,z)$ for the unknot 
vortex. The line where $\varphi$ jumps by $\pi$ is clearly visible.
The configuration is $z \to -z$ symmetric.

\vskip 0.4cm
{\bf Figure 3:} A combined surface and contour plot
of energy density $H(r,
z)$ for the unknot vortex, for comparison
as figure 1.
The configuration is $z \to - z$ symmetric. 

\vskip 0.4cm
{\bf Figure 4:} Three dimensional view of the extended
core for the unknot
vortex. The core is defined as the region
where energy density essentially vanishes.

\vskip 0.4cm
{\bf Figure 5:} The same as figure 5 but for a trefoil with
$g^\star \approx 1.3, \ e = 1$. 
The center denotes the core of the initial 
configuration, determined using a minimum energy 
principle \cite{simon}.
We thank J. Simon for providing us with the initial 
parametrization.

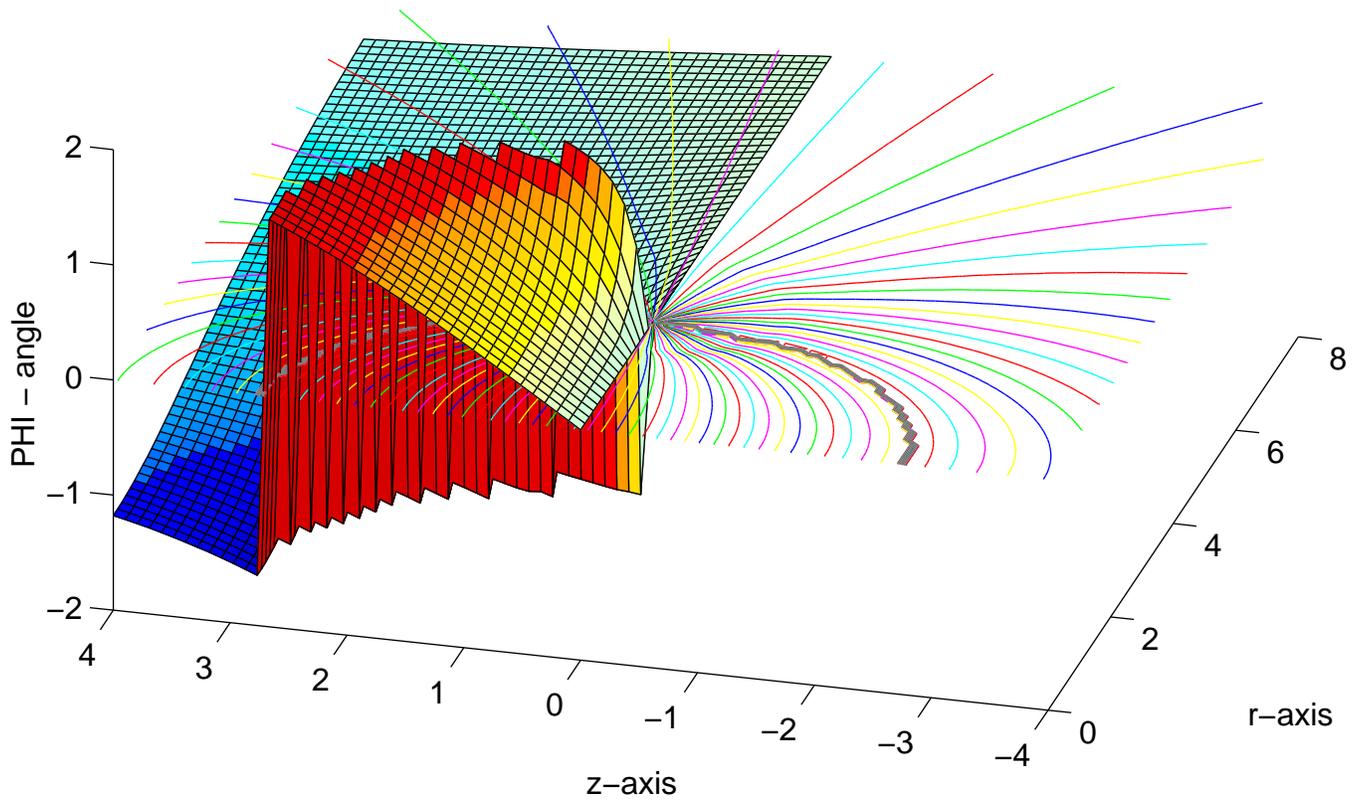
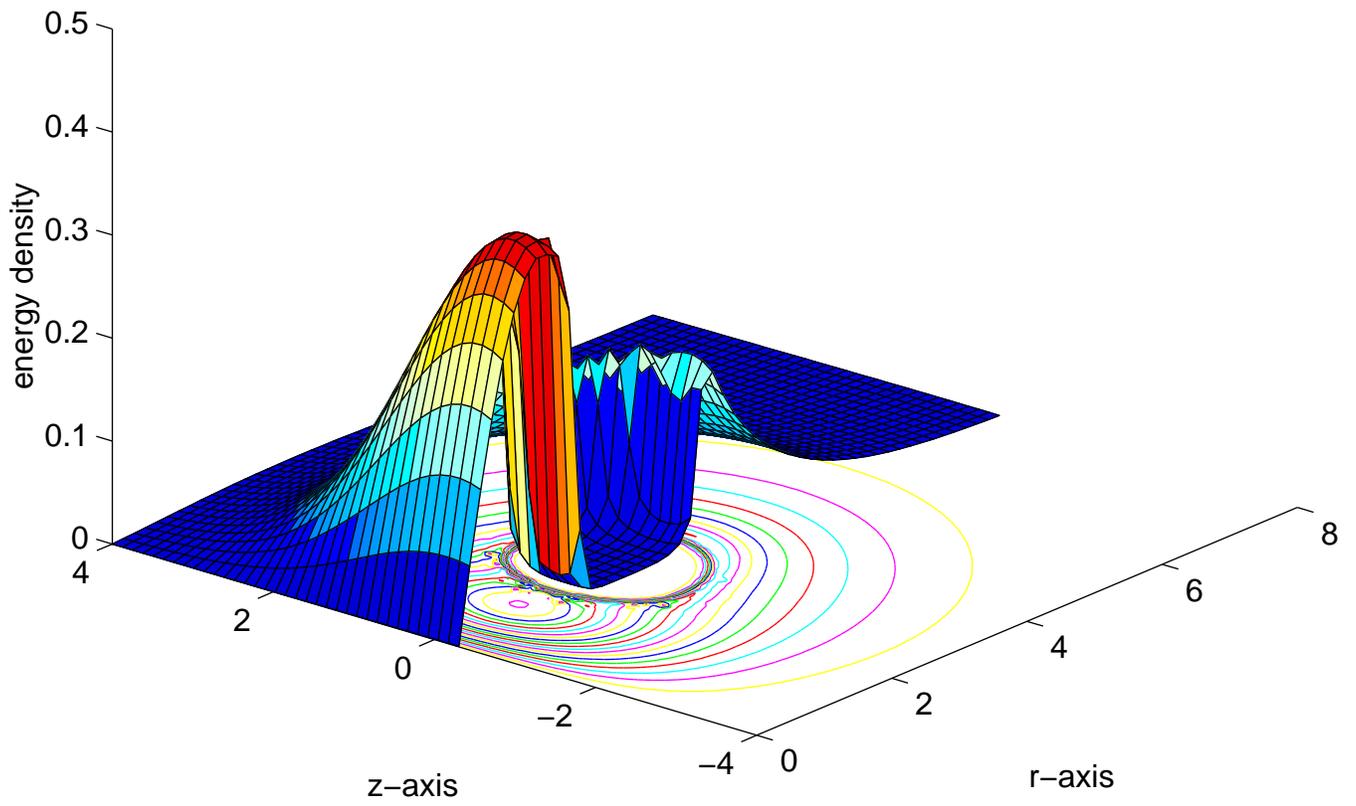
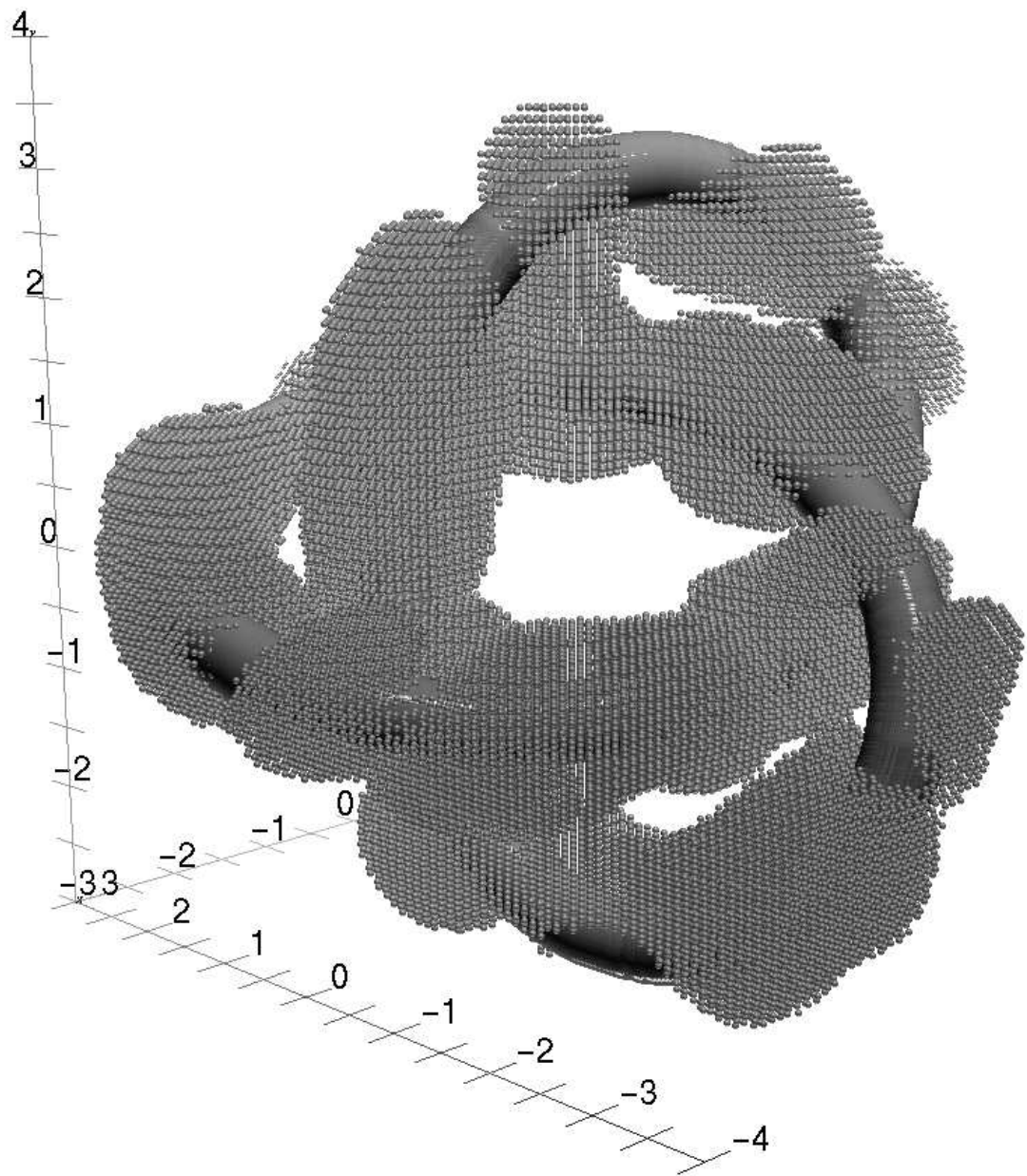
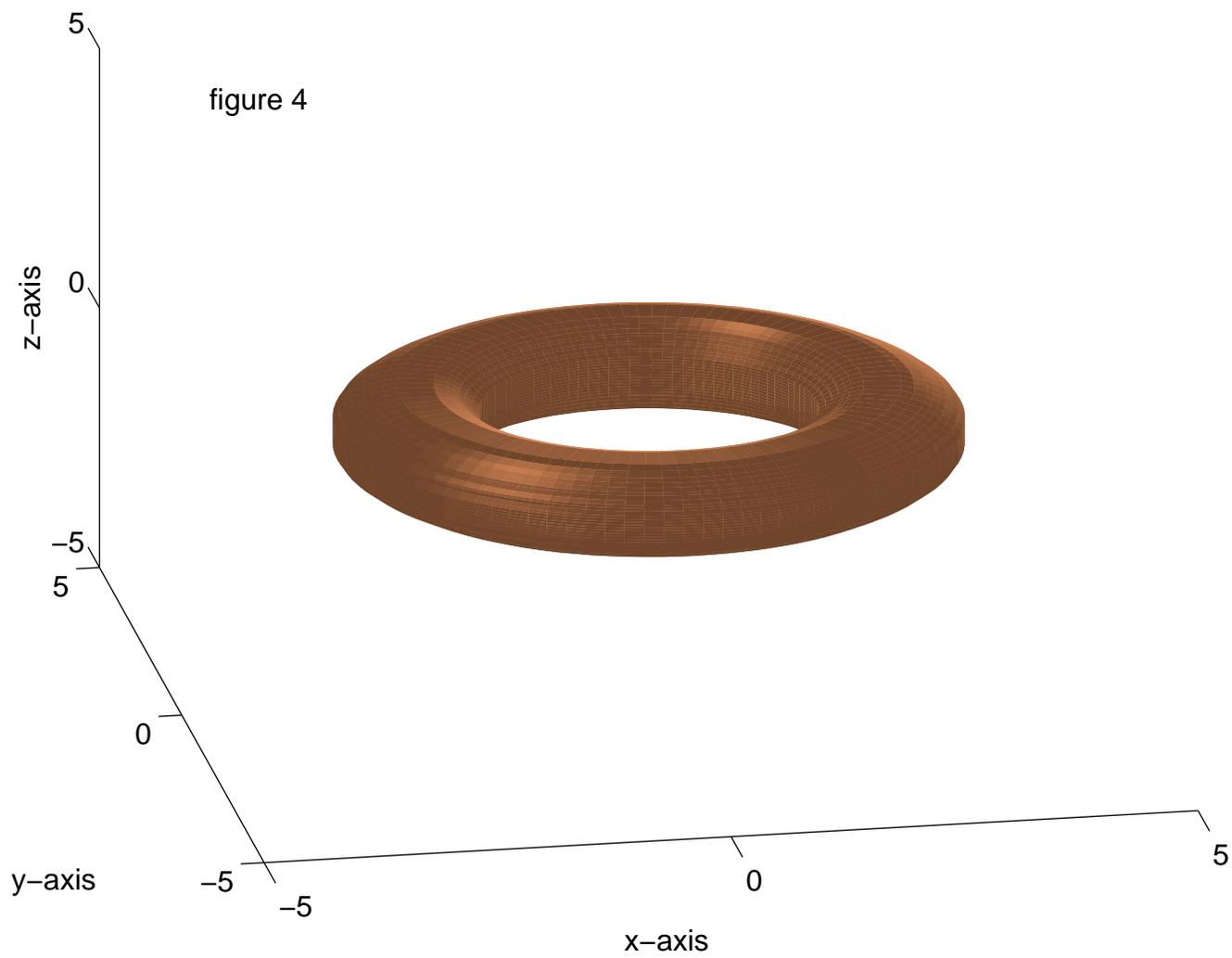
\end{document}